# Effect of High-κ Dielectric Layer on 1/$f$ Noise Behavior in Graphene Field-Effect Transistors

Yifei Wang,[1,‡] Vinh X. Ho,[1,‡] Zachary. N. Henschel,[1] Michael P. Cooney,[2] and Nguyen Q. Vinh[1,*]

[1] Department of Physics and Center for Soft Matter and Biological Physics, Virginia Tech, Blacksburg, VA 24061, USA
[2] NASA Langley Research Center, Hampton, Virginia 23681, USA

[‡]Yifei Wang and Vinh X. Ho contributed equally to this work.
[*] Corresponding author: vinh@vt.edu; phone: 1-540-231-3158

**Keywords:** Graphene, field-effect transistor, flicker noise, 1/$f$ noise, high-κ dielectric thin film, atomic layer deposition

**ABSTRACT**

We report the 1/$f$ noise characteristics at low-frequency in graphene field-effect transistors that utilized a high-κ dielectric tantalum oxide encapsulated layer (a few nanometers thick) placed by atomic layer deposition on Si$_3$N$_4$. A low-noise level of ~ $2.2 \times 10^{-10}$ Hz$^{-1}$ has been obtained at $f$ = 10 Hz. The origin and physical mechanism of the noise can be interpreted by the McWhorter context, where fluctuations in the carrier number contribute dominantly to the low-frequency noise. Optimizing fabrication processes reduced the number of charged impurities in the graphene field-effect transistors. The study has provided insights into the underlying physical mechanisms of the noise at low-frequency for reducing the noise in graphene-based devices.

**INTRODUCTION**

The outstanding electrical properties of graphene (a single atomic layer) have received considerable attention for future electronics including high speed transistors, photodetectors, components of integrated circuits, flexible and wearable devices, touch screens, ultrasensitive sensors.[1-3] In these applications, the flicker or low-frequency 1/$f$ noise ($f$ < 100 kHz) is the key factor of the device performance. The amplitude of the flicker noise defines the limit of the operation of electronics devices.[4-8] Therefore, to enhance the performance of graphene devices, several configurations have been realized to scale down the flicker noise. Graphene field-effect transistors (GFETs) containing a few graphene layers helped reduce noise level.[9] GFETs with the graphene channel on h-BN or encapsulated by two h-BN layers, which reduce charged impurities and trapping sites, can suppress the flicker noise by a factor of 5 to 10 times compared to that of graphene on Si/SiO$_2$ substrates.[10-12] The noise at low-frequency of graphene devices can be reduced through irradiation.[5, 13] Another approach is to employ a high-κ dielectric layer that can protect graphene from exposure conditions to prevent the increase of the noise. Nevertheless, the mobility degradation in graphene can occur when the dielectric layer is grown atop graphene. The flicker noise in GFET devices with an HfO$_2$ high-κ dielectric thin film grown by atomic layer deposition (ALD) has been realized.[14] A reduction of the low-frequency noise with a top-gated GFET using Al$_2$O$_3$ as gate-dielectric has been reported.[15] However, a high dark current in the milliampere scale of such GFETs is an obstruction for many applications. For example, the high dark current in GFET photodetectors causes a high shot noise, and thus, sets a high-level noise floor of these devices. These reports mostly focused on GFETs under the voltage control of the top-gate. However, the back-gated FETs are the backbone for a variety of electronic applications. Therefore, it is essential to improve GFETs with low noise in the back-gated configuration, which is covered by a high-κ dielectric thin film with a few nanometers thick. Among various high-κ dielectric materials, we have focused on the tantalum pentoxide (Ta$_2$O$_5$) with high dielectric constant (κ = 25 – 40) and good chemical and thermal stability. The material has been used in many applications in solar energy conversion as well as microelectronics, including photocatalytic materials,[16] charge-trapping for nonvolatile resistive random access memories,[17] atomic switches,[18] capacitors, insulators,[19] thin-film electroluminescent devices,[20] and high-speed elements.[21]

Here, we report the reduction of the flicker noise in GFETs by engineering the high-quality dielectric tantalum oxide ($Ta_2O_5$) layer grown by ALD on $Si_3N_4$. The back-gate bias dependence of the flicker noise on the graphene channel size with the source-drain distance varied from 10 to 200 µm has been investigated systematically. The noise magnitude has been observed to be a factor of 10 times lower in comparison with that in recent reports. The normalized noise-power spectral density of ~ $2.2 \times 10^{-10}$ $Hz^{-1}$ at $f$ = 10 Hz has been obtained. The noise mechanism can be explained by fluctuations of the carrier number, which are originated from the carrier trapping and de-trapping processes.

**RESULTS AND DISCUSSION**

To investigate the flicker noise, we fabricated GFETs on different substrates (Si/$SiO_2$ and Si/$Si_3N_4$) with and without an encapsulated high-κ dielectric layer ($Ta_2O_5$). A highly *p*-doped Si wafer (1 – 10 Ω.cm) and a 300-nm $SiO_2$ (or 300-nm $Si_3N_4$) layer are employed as the back-gate and dielectric layer, respectively. Graphene transistors were prepared in following steps. First, photolithography, electron-beam deposition and lift-off processes were employed to form metal contacts of Cr (3 nm) and Au (100 nm) for source, drain and back-gate. Second, a single graphene sheet was transferred onto the $SiO_2$/Si or $Si_3N_4$/Si substrate. A $Ta_2O_5$ dielectric layer has been grown on top of graphene for some GFET devices in two steps, including growing a 2-nm $Ta_2O_5$ seed layer by electron beam evaporation and an 18-nm $Ta_2O_5$ film by ALD.[22-23] Photolithography and dry etching steps were employed to determine the active size of the devices. The distances between source and drain (length), $L$, are 10, 20, 30, 50, 75, 100, 150, 200 µm, and the ratio between the width and length, $W/L$, of the active area is fixed at 2. The detail of the fabrication is provided in the Supplementary Information. A diagram of fabrication steps is illustrated in Figure S1. A surface image of a graphene sheet on $Si_3N_4$/Si using atomic force microscope (AFM) is provided in Figure S2.

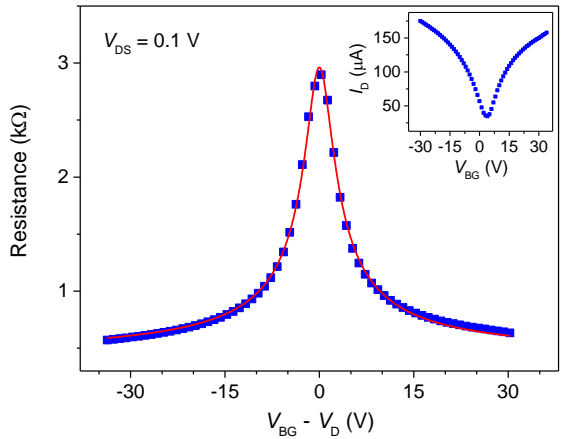

**Figure 1**: Resistance – voltage transfer characteristics of a graphene device covered by an ALD $Ta_2O_5$ film on $Si_3N_4$ ($L$ = 10 µm, $W$ = 20 µm) at room temperature under $V_{DS}$ = 0.1 V. The left inset provides a schematic of our graphene device, and the right inset shows the original current – voltage transfer curve.

The current – voltage (I-V) behavior was characterized by two Keithley source-meters units. A Keithley 2400 was employed to vary the back-gate voltage, $V_{BG}$, while a Keithley 2450 was used to set a constant voltage between drain and source contacts, $V_{DS}$, and to measure the drain current, $I_D$. Our electrical measurements were performed at room temperature. The I-V characteristics of a GFET covered by an ALD $Ta_2O_5$ film on $Si_3N_4$ ($L$ = 10 µm, $W$ = 20 µm) under $V_{DS}$ = 0.1 V is provided as an example in Figure 1, inset. Details of the setup and I-V characteristics at different $V_{DS}$ are illustrated in Figures S3 and S4.

To accurately determine the mobility, $\mu$, of carriers in graphene, the contact resistance, $R_c$, on a level with the graphene channel resistance, $R_{ch}$, is estimated from the total device resistance, $R = V_{DS}/I_D$. The mobility of carriers in the graphene device can be extracted by fitting the resistance – voltage (R-V) characteristic curve (Figure 1) in the following form.[24-26]

$$R = 2R_c + R_{ch} = 2R_c + \frac{L}{Wq\mu}\frac{1}{\sqrt{n_0^2 + n_g^2}}, \quad (1)$$

where $q$ is the charge of electron, $n_0$ is the carrier density resulting from charged impurities at the interface between dielectric layers and graphene or in the dielectric layers, $n_g = \frac{C_G}{q}(V_{BG} - V_D)$ is the density of charged carriers generated by a voltage on the back-gate away from the charge neutrality point voltage, $V_D$, (the Dirac point), $C_G = \frac{\varepsilon\varepsilon_0}{d}$ is the areal capacitance of the gate, with $d$ being the thickness of the $Si_3N_4$ dielectric material ($d$ = 300 nm), $\varepsilon_0$ is the vacuum permittivity, and $\varepsilon \sim 6$ is the dielectric constant of $Si_3N_4$.[27] The capacitance of the 300-nm $Si_3N_4$ dielectric layer is ~17 nF/cm$^2$. The red curve in Figure 1 presents the best fit to the R-V characteristic data using Eq. 1. We obtained the contact resistance at each electrode of ~209 Ω. The carrier density generated by charged impurities is ~2.4 × 10$^{11}$ cm$^{-2}$ at Dirac point. A high carrier mobility has been obtained in the GFET of ~5080 cm$^2$/(V·s).

The level of the flicker noise is the key point for the performance of graphene devices. The 1/$f$ noise was identified using a 100 kHz FFT spectrum analyzer (SR770) at room temperature. A current amplifier (FEMTO DLPCA-200) was employed to amplify the drain current. At a fixed back-gate voltage, the drain current was varied by adjusting drain-source voltage, $V_{DS}$, from 20 mV to 0.5 V. The noise-power spectral density can be described as,

$$S_I = \frac{AI_D^2}{f^\gamma}, \quad (2)$$

where $A$ is the amplitude of the noise and follows the empirical relation, $A = \alpha_H/n$, with $\alpha_H$ being the Hooge's noise parameter to evaluate the magnitude of the flicker noise,[4-5] $n$ is the total amount of carriers passing through the conducting area, $\gamma \approx 1$ is an experimental value.[5,7]

The noise-power spectral density of GFETs on $Si_3N_4$ ($L \times W$ = 10 × 20 µm$^2$) with the ALD $Ta_2O_5$ encapsulated layer was investigated under different drain current from 6.18 to 156.8 µA, and the voltage of the back-gate was fixed at Dirac point ($V_{BG} = V_D$ = 1 V). A typical noise-power spectral density under $V_{DS}$ = 0.1 V is shown in Figure 2a (upper inset). This plot follows the 1/$f$ dependence (namely, $\gamma$ = 1), and the noise-power spectral density, $S_I$, is ~10$^{-18}$ A$^2$ Hz$^{-1}$ at 1 Hz. The absence of bulges on the noise-power spectral density in this GFET device indicates that traps with a specific time constant are not dominated in the spectrum. Note that a few bulges have been reported in GFETs on Si/SiO$_2$. The bulges come from the generation-recombination (G-R) noise indicating fluctuation processes with well-defined frequencies.[5] When the drain current, $I_D$, was varied from 6.18 to 156.8 µA, the noise-power spectral density increases with a square function from 3.2 × 10$^{-19}$ to 7.7 × 10$^{-17}$ A$^2$ Hz$^{-1}$ at $f$ = 1 Hz as plotted on a linear scale (Figure 2a) and a log-log plot (Figure 2a, lower inset). The noise-power spectral density proportional to $I_D^2$ is also reported in previous reports for graphene devices on SiO$_2$.[5,10]

To evaluate the noise of GFETs on different insulator/dielectric layers, we determine the normalized noise-power spectral density, in which the noise is normalized with current, $S_I/I_D^2$, or the noise behavior independent of the drain current flowing through the device. The experiments performed under $V_{DS}$ = 0.1 V and $V_{BG} = V_D$ = 1 V for graphene devices on SiO$_2$, $Si_3N_4$, and on $Si_3N_4$ with the ALD $Ta_2O_5$ encapsulated layer (Figure 2b). These devices were fabricated under the same conditions and graphene channel size of $L$ = 10 µm and $W$ = 20 µm. The noise level at $f$ = 1 Hz of the graphene device on SiO$_2$ without the ALD

Ta$_2$O$_5$ layer is $1.8 \times 10^{-8}$ Hz$^{-1}$, which is similar to previous reports.[10, 12] The graphene on Si$_3$N$_4$ device without the ALD Ta$_2$O$_5$ layer shows a lower noise level of $0.8 \times 10^{-8}$ Hz$^{-1}$, which is about 2 times lower than that of graphene on SiO$_2$ device. However, the graphene device covered with the high-κ Ta$_2$O$_5$ dielectric layer on Si$_3$N$_4$ shows a noise level of $2.5 \times 10^{-9}$ Hz$^{-1}$. At $f = 10$ Hz, the normalized noise-power spectral density, $S_I/I_D^2$, of this device is equal to $2.2 \times 10^{-10}$ Hz$^{-1}$, indicating a reduction of the 1/$f$ noise by 5 and 50 times as compared to that of graphene devices on h-BN and SiO$_2$, respectively.[10-12]

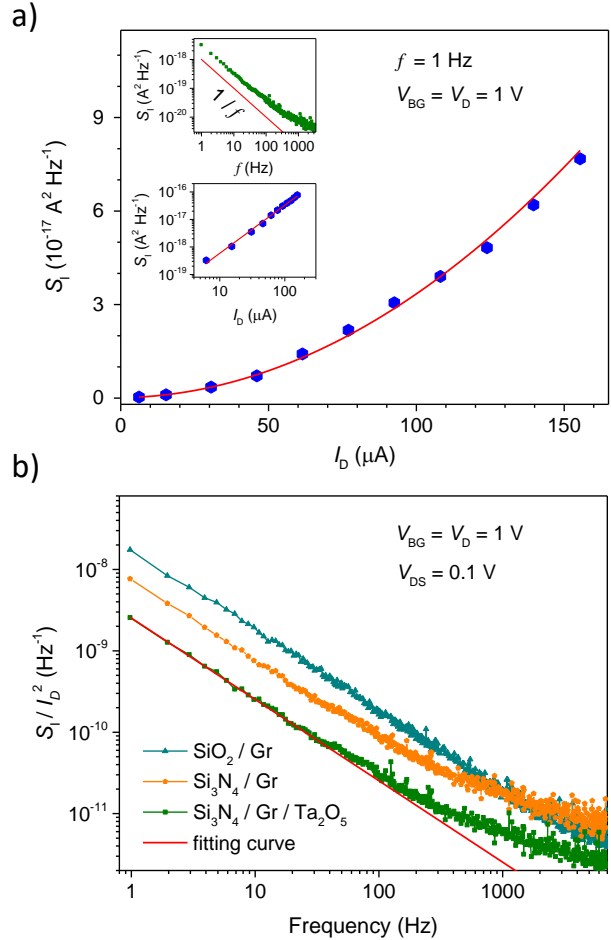

**Figure 2:** The flicker noise in GFETs ($L = 10$ μm, $W = 20$ μm) at Dirac point voltage (mainly, $V_{BG} = V_D = 1$) and under $V_{DS} = 0.1$ V. (**a**) The noise-power spectral density at $f = 1$ Hz increases with a square function of the drain current. Upper inset provides the noise-power spectral density, while lower inset presents the noise-power spectral density at $f = 1$ Hz. (**b**) The noise-power spectral density is normalized with current in graphene on SiO$_2$, Si$_3$N$_4$, and on Si$_3$N$_4$ with the ALD Ta$_2$O$_5$ encapsulated layer.

The gate-bias characteristics of the noise-power spectral density normalized with current were further examined for the graphene device on Si$_3$N$_4$ with the ALD Ta$_2$O$_5$ layer under different back-gate voltage. The normalized noise-power spectral density, $S_I/I_D^2$, and together with the amplitude of the noise, $A = (1/M) \sum_{m=1}^{M} f_m S_{Im}/I_m^2$, as function of gate voltage, ($V_{BG} - V_D$), for graphene devices on SiO$_2$, Si$_3$N$_4$, and on Si$_3$N$_4$ with the ALD Ta$_2$O$_5$ encapsulated layer at $f = 1$ Hz are illustrated in Figures 3 and S5, respectively. Note that the noise amplitude is an average over several frequencies, thus, is the same as the noise-power spectral density normalized with current, $S_I/I_D^2$. As shown in Figure 3, M-shape behaviors of the noise-power spectral density normalized with current have been observed in devices with graphene on $SiO_2$ and $Si_3N_4$ with local minimum at $V_{BG} = V_D$. The observation of the noise behavior agrees with previous reports,[10, 12, 28] indicating that the noise at low-frequency in our devices on $SiO_2$ is typical for graphene. The characteristic is related to the presence of the spatial charge inhomogeneity as well as electron-hole puddles in graphene devices on $SiO_2$.[28] In contrast, the behavior is not observed in graphene devices on $Si_3N_4$ with the ALD $Ta_2O_5$ layer. The noise-power spectral density normalized with current together with the amplitude of the noise are more than one order of magnitude lower in the graphene device on $Si_3N_4$ with the ALD $Ta_2O_5$ encapsulated layer compared to that on $SiO_2$. A Λ-shape instead of an M-shape has been observed in the device (Figure 3). The noise level is lower when voltage applied to the back-gate is further away from Dirac point voltage. Graphene devices without a dielectric layer ($Ta_2O_5$) can absorb water vapor or organic contaminations under environmental exposures, which leads to a high noise level.[12, 29-30]

We now address the origin and physical mechanism of the flicker noise behavior in our devices. From a relation between the drain current, mobility and the number of charge carriers ($I_D \propto q\mu n$) in Eq. 1, fluctuations in the drain current can be described as $\delta I_D \propto qn\delta\mu + q\mu\delta n$.[5-6] The flicker noise in FET devices typically originates from fluctuations in the carrier mobility or the number of charge carriers, or both.[31] In the context of the fluctuations in the carrier mobility described by Hooge model, the carrier mobility in graphene is typically affected by long-range Coulomb scatterings associated with charged impurities, and short-range disorder scatterings related to intrinsic defects, cracks, or boundaries of graphene.[29-30] Other scattering processes including ripples, phonons, mid-gap states generate fluctuations in the carrier mobility similar to that expected from the long- and short-range scatterings.[1, 32-36] Thus, we will analyze the fluctuations in these two scattering sources separately. The carrier mobility limited by the long-range Coulomb scattering does not exhibit gate-bias dependence, whereas the mobility associated with the other scattering depends on the gate-bias.[1] The behavior of the noise will be determined by which scattering mechanism contributes dominantly. In the viewpoint of fluctuations in the carrier number expected from the McWhorter's relation, the fluctuations are induced by the number of charge carriers passing through graphene layer due to the trapping/de-trapping processes near graphene-dielectric interfaces. The efficiency of these processes is defined by the density of empty as well filled states near the Fermi level.

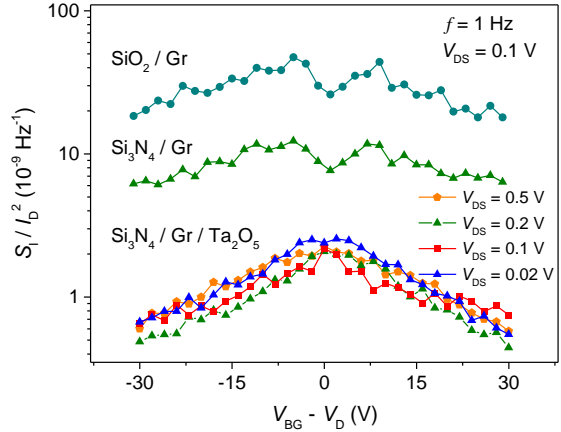

**Figure 3:** Normalized noise-power spectral density as function of back-gate voltage ($V_{BG}$ - $V_D$) for graphene on $SiO_2$, $Si_3N_4$, and $Si_3N_4$ with the ALD $Ta_2O_5$ encapsulated layer at $f$ = 1 Hz.

Our observation of an M-shape behavior in graphene on $SiO_2$ and $Si_3N_4$, devices (Figure 3) is consistent with previous reports of non-encapsulated GFET structures on $SiO_2$.[10, 29-30] In these structures, molecules from the air (i.e., water-like contaminants) trapped on graphene are likely adding to the source of the charge-density inhomogeneity (or charge puddles) near Dirac point, which arises the long-range Coulomb scattering across the graphene channel.[35-36] Transforming from the M- to V-shape after annealing these devices was assigned to the suppression of the long-range scattering by removing absorbed molecules, thus, reducing the fluctuations in the carrier mobility.[29] On the other hand, the M-shape noise behavior can also originate from the high trap density in the device fabrication process.[28] These traps cause trapping/de-trapping processes of charged carriers near the graphene - SiO$_2$ interface that induces the fluctuations in number carrier.

The unexpected gate-bias behavior of the flicker noise in graphene on SiO$_2$ reported in several observations complies with the Hooge approach.[37] The measured noise can be interpreted by an empirical relation between the Hooge's noise parameter, $\alpha_H$, the mobility, μ, and a network of resistors.[30] While carrier mobility associated with the long-range Coulomb scattering is independent of the gate-voltage (i.e., $\mu_L = 1/C_L$), the carrier mobility related to the short-range scattering is inversely proportional to voltage applied on the back gate, $\mu_S = 1/(C_S(V_{BG} - V_D))$, where $C_L$ and $C_S$ are the long- and short-range scattering constants.[38-39] Following the dependence of the Hooge parameter (i.e., $\alpha_H \approx (1/\mu)^\delta$) on the carrier mobility[30] and Matthiesen's rule (i.e., $1/\mu = 1/\mu_L + 1/\mu_S$), the dependence on the gate bias of normalized noise-power spectral density is expressed as,[30]

$$\frac{S_I}{I^2} = \frac{\alpha_H}{nf^\gamma} \approx \frac{(C_S(V_{BG}-V_D)+C_L)^\delta}{(V_{BG}-V_D)f^\gamma}, \tag{3}$$

where the $\delta \approx 3$ is experimentally determined for graphene on a substrate device.[30, 40] The 1/$f$ noise is contributed by both the short- and long-range carrier scattering. The transport property of carriers near Dirac point in graphene channel is governed by spatial charge inhomogeneity which is related to the presence of electron-hole puddles in graphene, resulting in an increase in fluctuations.[28] The V-shape have been observed. For voltage applied on the back-gate away from Dirac point, fluctuations are governed by long-range Coulomb scattering, thus, the 1/$f$ noise decreases with increasing $V_{BG}$.[30] The M-shape behavior has been observed in our graphene device on SiO$_2$.

To improve the device performance as well as to reduce the flicker noise in our graphene devices, we have employed a high-κ dielectric material, Si$_3$N$_4$, for the insulator gate, and encapsulated graphene with an ALD Ta$_2$O$_5$ layer. As mentioned above, the normalized noise-power spectral density is more than one order of magnitude lower in the graphene device on Si$_3$N$_4$ with the ALD top layer as compared to that on SiO$_2$. The observed 1/$f$ noise behavior in these devices presents the Λ-shape dependence (Figure 3). The short-range disorder scattering does not primarily attribute to the flicker noise of our devices. Therefore, the carrier-mobility fluctuations in the short- and long-range scatterings associated with the Hooge model do not contribute dominantly to the flicker noise, whereas the fluctuations in the carrier number associated with the McWhorter approach have a major role in the flicker noise of these devices.

The decrease of impurities in graphene devices on Si$_3$N$_4$ with an ALD Ta$_2$O$_5$ encapsulated layer is expected from our fabrication process. Although our devices are not annealed to remove atmospheric contaminants, the 1/$f$ noise reduction and Λ-shape behavior have been obtained. Typically, metal contacts were implemented by depositing metals on top of graphene on a substrate. This method caused the contamination of photoresist residues on graphene during the photolithography process. In contrast, our fabrication process starts with a well-prepared structure of Si/Si$_3$N$_4$/metal-contacts, and a single graphene sheet is transferred onto this structure. Graphene surface is not covered by polymer at any stage of the fabrication, reducing the number of traps at interfaces of graphene/metal contacts and graphene/Si$_3$N$_4$. The reduction of photoresist residuals on graphene surface before growing the ALD Ta$_2$O$_5$ layer was confirmed by AFM image (Supplementary information, Figure S2).

In order to reduce the fluctuations in the carrier mobility, the high-κ ALD Ta$_2$O$_5$ dielectric layer has been deposited on the top of our graphene devices to suppress the long-range Coulomb scattering as well as protect the graphene surface. Although a degradation of the carrier mobility in graphene could happen when the dielectric layer was grown on top of graphene, this dielectric layer with high quality could protect graphene from exposure conditions to prevent the increase of 1/$f$ noise.[14] The dielectric screening effect of the high-κ Ta$_2$O$_5$ dielectric material can minimize the long-range Coulomb scattering efficiently.[41] On the other hand, to achieve a uniform deposition of the ALD layer on graphene, a 2-nm

Ta$_2$O$_5$ seed layer was deposited by electron beam evaporation. The samples were loaded into a vacuum chamber (3 × 10$^{-6}$ Torr), which significant reduced contaminants on the surface of graphene before growing the 2-nm Ta$_2$O$_5$ seed layer. During the evaporation step, the 2-nm Ta$_2$O$_5$ layer contains oxygen vacancies, acting as carrier traps at the interface between graphene and the seed layer. The oxygen vacancy-related traps can generate fluctuations in the carrier number through random trapping/de-trapping processes of carriers. Thanks to H$_2$O pulses during the first few cycles of the ALD process at 300 °C, the 2-nm TaO$_x$ seed layer is fully oxidized, resulting in a low number of carrier traps in the seed layer. Consequently, the suppression of the long-range scattering as well as fewer effective trap states at the interfaces between graphene and oxide resulted in the reducing of the noise and produced the Λ-shape gate-bias behavior. Note that by varying the voltage applied on the gate, we can control the carrier number in the active area. As mentioned before, the normalized noise-power spectral density does not follow the simple $1/(V_{BG} - V_D)$ dependence as expected from the Hooge approach (Figure 3). The Λ-shape gate-bias behavior suggests that another physical mechanism for the flicker noise of the drain current dominates in our GFETs.

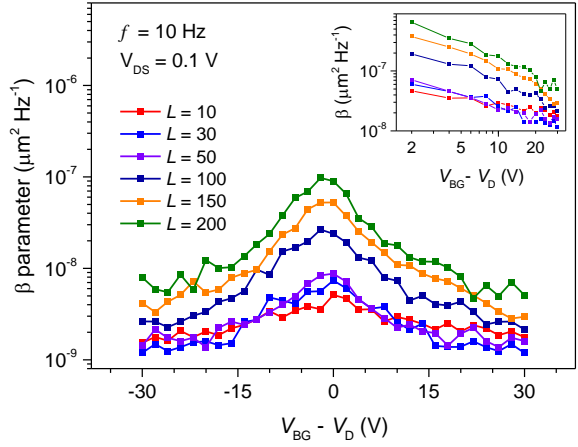

**Figure 4:** Area-normalized noise-power spectral density, $\beta = (S_I/I_D^2)(L \times W)$, at $f = 10$ Hz plots again the back-gate voltage ($V_{BG} - V_D$) for graphene devices on Si$_3$N$_4$ with the ALD Ta$_2$O$_5$ encapsulated layer. Inset shows the data on the log-log scale.

To shed light on the mechanism of the flicker noise, we further investigated noise characteristics with different device area. The active area, $L \times W$, is varied in a wide range from 200 to 80000 µm$^2$, where the lengths, $L$, are 10, 20, 30, 50, 75, 100, 150, 200 µm, and the ratio of $W/L$ is fixed at 2. The voltage between source and drain, $V_{DS}$, was fixed at 0.1 V. For the device with $L \times W = 200 \times 400$ µm$^2$, at Dirac point ($V_{BG} = V_D$) and $f = 10$ Hz, a maximum value of ~ 1 × 10$^{-12}$ Hz$^{-1}$ has been observed for the normalized noise-power spectral density (Figure S6). The value is several orders of magnitude lower than that in devices on SiO$_2$ reported in the literature.[10-12] To compare the flicker noise behavior in the graphene devices with different active area, we employ the area-normalized noise-power spectral density, $\beta = (S_I/I_D^2)(L \times W)$.[10-11] Figure 4 shows plots of $\beta$ parameter at $f = 10$ Hz, ranging between 1 × 10$^{-9}$ to 5 × 10$^{-9}$ µm$^2$ Hz$^{-1}$ for the length, $L$, of < 50 µm. In the micrometer scale, the flicker noise in these devices was reduced by a factor of 5 as compared to that in graphene devices encapsulated by two h-BN layers,[11] a HfO$_2$ dielectric gate[14], or on h-BN[10] with $\beta$ reported from 5 × 10$^{-9}$ to 10$^{-7}$ µm$^2$ Hz$^{-1}$ at $f = 10$ Hz. The value is the same order with multiple graphene layers on a substrate.[42] However, the mobility of carrier decreases with adding graphene layers,[42-43] thus, reducing the performance of the devices. Note that under identical conditions including gate-bias and temperature, the carrier number is proportional to the active area.[12] As illustrated in Figure 4, the area-normalized noise-power spectral density does not scale inversely proportional to the device area (namely, the simple $1/n$ dependence or the graphene sheet resistant).[44]

This suggests that the reduction of the noise and the Λ-shape behavior of the devices covered with the ALD $Ta_2O_5$ dielectric layer cannot be explained with the Hooge's model.

To provide insight into the Λ-shape behavior, we employed the framework of the McWhorter approach based on the fluctuations in the carrier number of single-layer graphene in field-effect transistors. The amount of charged carriers in the active area can be varied by sweeping the back-gate voltage further away from Dirac point, and this value is proportional to the gate voltage, $V_{BG}$ (see Eq. 1).[24] In the McWhorter model, normalized noise-power spectral density, $S_I/I_D^2$, reduces with the $1/n^2$ dependence, thus, the $β$ parameter is proportion to $1/n$ or $1/V_{BG}$.[12] The carrier tunneling to/from the graphene channel through the trapping/de-trapping processes gives rise to the fluctuations in the carrier number in the graphene channel, which mainly contributes to the $1/f$ noise. The inset of Figure 4 shows the $β$ parameter is scaled down with the back-gate bias, $V_{BG}$. The noise characteristic in our devices can be explained by the McWhorter model, in which the fluctuations in the carrier number are the dominant source for the flicker noise.

Following the McWhorter approach, we can estimate the effective trap density, $D_{eff}$, at Fermi level from normalized noise-power spectral density:[45]

$$\frac{S_I}{I^2} = \frac{k_B T D_{eff}}{fLWn_c^2 \ln(\tau_{max}/\tau_{min})}, \quad (4)$$

where $T$ is the temperature, $\tau_{min}$ and $\tau_{max}$ are the minimum and maximum time for carrier tunneling, respectively, and $k_B$ is Boltzmann's constant. The carrier concentration in the active area can be estimated from the charged carrier density, $n_c = \sqrt{n_0^2 + n_g^2}$. The approach has been used to analyze the low-frequency noise in metal-oxide-semiconductor field-effect transistors based on Si or GaAs.[45] To estimate $D_{eff}$, we fit the experimental results to the McWhorter model (Eq. 4) with $\ln(\tau_{max}/\tau_{min}) = 4$,[15, 45] (the red line in Figure 2b). Under $V_{DS} = 0.1$ V, the effective trap density, $D_{eff}$, is ~$1.14 \times 10^{10}$ cm$^{-2}$ eV$^{-1}$ at Dirac point (i.e., $V_{BG} = V_D$), which is a factor of 10 times lower than that of GFETs on the $SiO_2$/Si substrate.[32-33, 46]

## CONCLUSIONS

In summary, we have carried out measurements of the low-frequency $1/f$ noise on graphene field-effect transistors covered by the high-$k$ dielectric $Ta_2O_5$ layer (a few nanometers thick) on $Si_3N_4$. A low noise level of ~ $2.2 \times 10^{-10}$ Hz$^{-1}$ has been obtained at $f = 10$ Hz. The dependence on the channel graphene area of the noise was also investigated systematically. The origin and physical mechanism of the noise can be interpreted by the McWhorter model, in which the tunneling of carriers from/to the graphene channel via the trapping/de-trapping processes in dielectric layers processes is the determining factor. The considerable suppression of flicker noise can offer guidance on practical implications.

## ASSOCIATED CONTENT

**Supporting Information**: Details for GFET fabrication, atomic force microscopy images, electrical setup, I-V transfer characteristics, noise amplitude, normalized noise-power spectral density.

## AUTHOR INFORMATION

**Corresponding Author:** *E-mail: vinh@vt.edu
**Author Contributions:** ‡Y. Wang and V. X. Ho contributed equally to this work.
**Notes:** The authors declare no competing financial interest.

## ACKNOWLEDGMENTS

The authors gratefully acknowledge the financial support of this effort by the Earth Science Technology Office (ESTO), NASA.

# SUPPORTING INFORMATION

# Effect of High-κ Dielectric Layer on 1/$f$ Noise Behavior in Graphene Field-Effect Transistors

Yifei Wang,[1‡] Vinh X. Ho,[1‡] Zachary. N. Henschel,[1] Michael P. Cooney,[2] and Nguyen Q. Vinh[1]*

[1] Department of Physics and Center for Soft Matter and Biological Physics, Virginia Tech, Blacksburg, VA 24061, USA
[2] NASA Langley Research Center, Hampton, Virginia 23681, USA

‡Yifei Wang and Vinh X. Ho contributed equally to this work.
* Corresponding author: vinh@vt.edu; phone: 1-540-231-3158

## 1. Device fabrication

Graphene field-effect transistors (GFETs) were fabricated on two different dielectric layers including silicon dioxide ($SiO_2$) or silicon nitride ($Si_3N_4$) on $p$-doped Si wafers (1 – 10 Ω.cm). A 300-nm $SiO_2$ layer grown thermally at 1050 °C includes a 200-nm wet thermal $SiO_2$ layer in between two 50-nm dry thermal $SiO_2$ layers. A 300-nm $Si_3N_4$ layer was grown by the plasma-enhanced chemical vapor deposition (PECVD) at 350 °C. Source and drain contacts were defined by photolithography, and deposited 3-nm Cr / 100-nm Au film by the e-beam evaporation method. To eliminate photoresist residue from the lift-off process, the Si/$SiO_2$ or Si/$Si_3N_4$ wafer with metal contacts was cleaned by oxygen plasma for 4 minutes to eliminate photoresist residue.

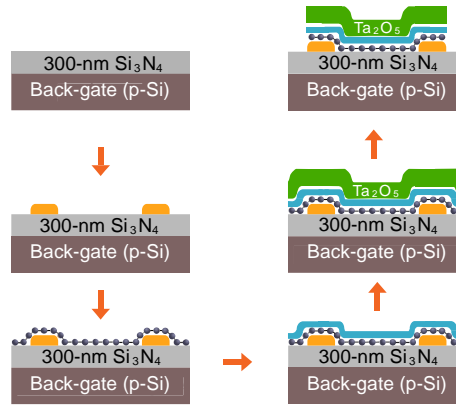

**Figure S1.** Schematic diagram of the GFET device fabrication.

Graphene was grown on copper (Cu) foil (18-μm) by chemical vapor deposition (CVD) from Graphenea Inc. The single-layer graphene films were confirmed by Raman spectroscopy. Poly(methyl methacrylate) (PMMA, MicroChem 495 PMMA A4 - (4% in Anisole) - 495,000 molecular weight) solution was spin-coated on graphene / Cu foil at 1700 rpm for 30 seconds and dried in a vacuum in 2 hours. The Cu foil of the CVD graphene / Cu film was removed by a 0.3 M ammonium persulfate (($NH_4)_2S_2O_8$, Sigma−Aldrich, ≥ 98%) aqueous solution at 25 °C, which enables to minimize residues compared to other $Fe(NO_3)_3$ and $FeCl_3$ solutions.[1-2] After two hours, the graphene layer was rinsed in deionized water for two times in 5 minutes each time to remove residual Cu etchant. The graphene sheet was picked up by a well-prepared substrate with electrodes. Then, the sample was put in a vacuum overnight to promote a good adhesion. Hereafter, the sample was heated at 135 °C in the air for 20 minutes to enable the flattening of the graphene film and produce stronger adhesion. After that, PMMA was washed by soaking them in acetone in 1 hour at 50 °C, followed by IPA in 30 minutes at room temperature. Photolithography and oxygen plasma etching were employed to fabricate a graphene pattern with different sizes. The distances between source and drain, $L$, of the channel were 10, 20, 30, 50, 75, 100, 150, 200 µm, and the ratio between the width and length, $W/L$, of the device was fixed at 2.

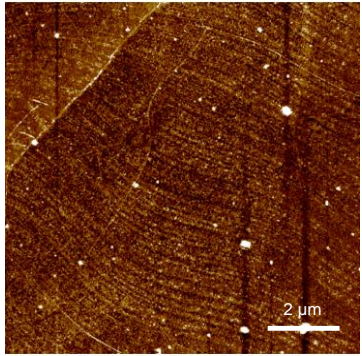

**Figure S2.** An AFM image of the graphene surface before growing the $Ta_2O_5$ film on GFET device. The size of the AFM image is $10 \times 10$ µm$^2$.

Next, the samples were loaded into an e-beam evaporation chamber (PRO line PVD 250, Kurt J. Lesker) and pumped down to $3 \times 10^{-6}$ Torr. Whereafter, a 2-nm $Ta_2O_5$ seed layer was grown with a rate of 0.1 Å/s. An 18-nm $Ta_2O_5$ film was then grown in an atomic layer deposition (ALD) system (Savannah S100 ALD, Cambridge Nanotech Inc.) at 300 °C. The precursors of pentakis(dimethylamino)tantalum (V) and water were sequentially exposed with nitrogen gas. The deposition rate is ~0.75 Å/cycle. A schematic diagram of the graphene-$Ta_2O_5$ heterostructure photodetector is illustrated in Fig. S1.[3-5]

Atomic force microscope (AFM) images have been performed to verify the quality of the graphene surface before growing the ALD $Ta_2O_5$ film. Figure S2 shows an AFM image of the graphene surface with a size of $10 \times 10$ µm$^2$. The graphene surface is clean, and a few white dots on the graphene surface are PMMA residues, thus most residuals were removed in our GFET devices.

## 2. Electrical measurements

To characterize the noise behavior, a Keithley 2400 unit was used to vary the back-gate voltage, $V_{BG}$, while a Keithley 2450 was employed to set a constant drain-source voltage, $V_{DS}$, and to measure the drain current, $I_D$. The drain-source bias was set between 20 mV to 0.5 V; however, most of the measurements were collected at 100 mV. A FEMTO DLPCA-200 low-noise current amplifier was employed to amplify the drain current. A 100 kHz FFT spectrum analyzer (Stanford Research 770 – SR770) with a high dynamics range of 90 dB has been used to characterize the noise behavior of graphene devices. The voltage noise-power spectral density, $S_V$, collected from the SR770 was converted into the current noise-power spectral density, $S_I$, by dividing the signal by the gain factor used in the current amplifier. The probe station is enclosed by a black aluminum anodized box to avoid ambient noise. This system is put on a vibration-isolation optical table to minimize the low-frequency vibration. All devices were measured after two weeks of fabrication.

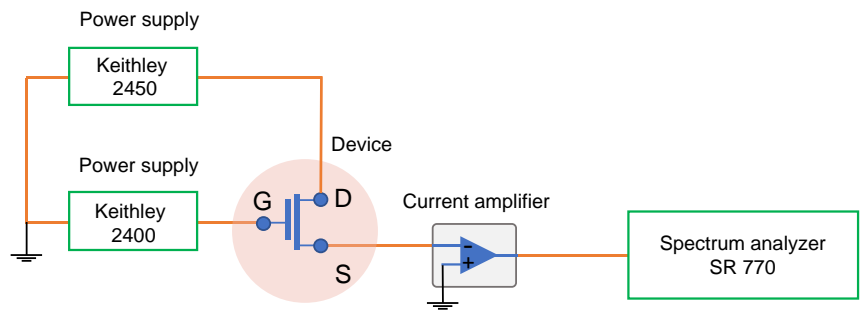

**Figure S3.** A schematic for electrical measurements.

## 3. Drain current under the different drain-to-source voltage

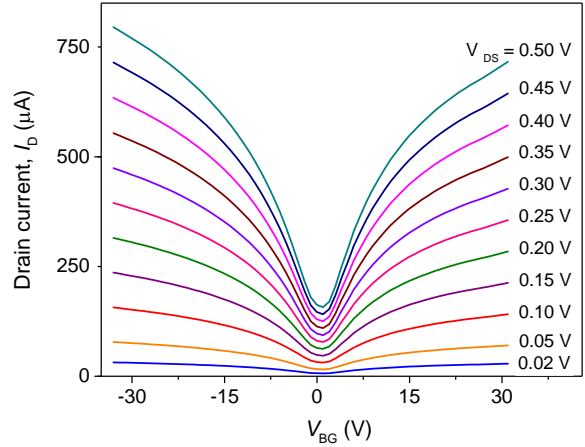

**Figure S4.** The current – voltage (I-V) characteristic curves of the GFET device on $Si_3N_4$ substrate ($L \times W = 10 \times 20$ µm$^2$) with the ALD $Ta_2O_5$ encapsulated layer.

## 4. Noise amplitude of the GFET device on $Si_3N_4$ substrate ($L \times W = 10 \times 20$ µm$^2$) with the ALD $Ta_2O_5$ encapsulated layer

To analyze the noise behaviors of GFETs with different size and dielectric layers, we calculate characteristic parameters, including the normalized noise-power spectral density, the noise amplitude as, the area-normalized noise-power spectral density, the area-normalized noise amplitude. Here is the list of these parameters:

1. The normalized noise-power spectral density, $S_I/I_D^2$, compares the noise level of graphene devices with different drain current values (Figures 2, 3, S6).

2. The noise amplitude, $A = (1/M)\sum_{m=1}^{M} f_m S_{Im}/I_m^2$, averages over different M frequencies for different drain currents passing through the device (Figures S5).

3. The area-normalized noise-power spectral density, $\beta = (S_I/I_D^2)(L \times W)$, is the noise characteristic independent of the size as well as the drain current passing through the device (Figures 4).

4. The area-normalized noise amplitude, $A(L \times W)$, is the noise amplitude normalized with different active area sizes of the device (Figure S7).

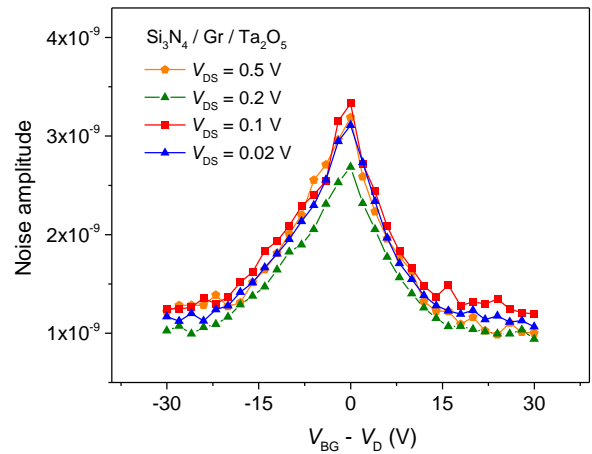

**Figure S5.** The noise amplitude behavior of the graphene device on $Si_3N_4$ with the ALD $Ta_2O_5$ encapsulated layer under different drain-source voltages. The size of the graphene channel is $L \times W = 10 \times 20$ µm$^2$.

## 5. The normalized noise-power spectral density of the 1/$f$ noise as function of the size of the GFET device on $Si_3N_4$ with the ALD $Ta_2O_5$ encapsulated layer

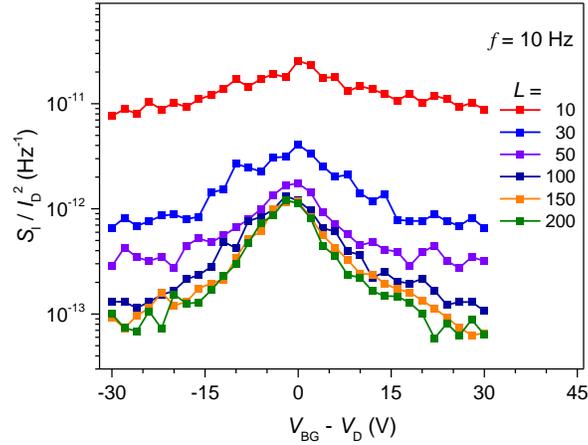

**Figure S6.** The normalized noise-power spectral density of GFETs on $Si_3N_4$ with the ALD $Ta_2O_5$ encapsulated layer with different graphene channel areas.

## 6. The dependence of the noise amplitude on the graphene channel area of the GFETs on $Si_3N_4$ with the ALD $Ta_2O_5$ encapsulated layer

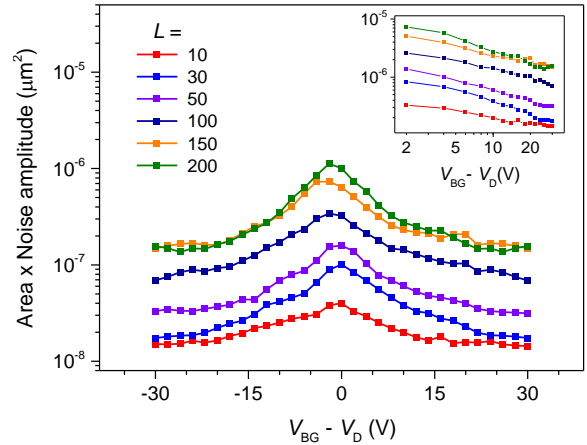

**Figure S7.** The dependence of the noise amplitude at $f$ = 10 Hz on the graphene channel area of GFET devices on $Si_3N_4$ with the ALD $Ta_2O_5$ encapsulated layer under $V_{DS}$ = 0.1 V. Inset shows the original data on the log-log scale.